\documentclass[sn-mathphys-num]{sn-jnl}

\usepackage{amsmath, amsfonts, amssymb}
\usepackage{graphicx}
\usepackage{booktabs}
\usepackage{microtype}
\usepackage{xspace}
\usepackage{url}
\usepackage{array}
\usepackage{tabularx}
\usepackage{siunitx}
\usepackage{hyperref}
\usepackage{orcidlink}

\newcommand{\sectionauthor}[1]{\begin{flushright}\textit{Contributed by #1}\end{flushright}\vspace{-1em}}

\title{A Survey of AI Methods for Geometry Preparation and Mesh Generation in Engineering Simulation}

\author*[1]{\fnm{Steven} \sur{Owen}\footnotemark[1]}\email{sjowen@sandia.gov}
\author[1]{\fnm{Nathan} \sur{Brown}\footnotemark[1]}
\author[2]{\fnm{Nikos} \sur{Chrisochoides}}
\author[3]{\fnm{Rao} \sur{Garimella}}
\author[4]{\fnm{Xianfeng} \sur{Gu}}
\author[5]{\fnm{Franck} \sur{Ledoux}}
\author[6]{\fnm{Na} \sur{Lei}}
\author[1]{\fnm{Roshan} \sur{Quadros}\footnotemark[1]}
\author[3]{\fnm{Navamita} \sur{Ray}}
\author[1]{\fnm{Nick} \sur{Winovich}\footnotemark[1]}
\author[7]{\fnm{Yongjie Jessica} \sur{Zhang}}

\affil[1]{\orgname{Sandia National Laboratories}, \orgaddress{\city{Albuquerque}, \state{NM}, \country{USA}}}
\affil[2]{\orgname{Old Dominion University}, \orgaddress{\city{Norfolk}, \state{VA}, \country{USA}}}
\affil[3]{\orgname{Los Alamos National Laboratory}, \orgaddress{\city{Los Alamos}, \state{NM}, \country{USA}}}
\affil[4]{\orgname{New York University / Stony Brook University}, \orgaddress{\state{NY}, \country{USA}}}
\affil[5]{\orgname{CEA, DAM, DIF, F-91297 Arpajon}, \orgaddress{\country{France}}}
\affil[6]{\orgname{Dalian University of Technology}, \orgaddress{\country{China}}}
\affil[7]{\orgname{Carnegie Mellon University}, \orgaddress{\city{Pittsburgh}, \state{PA}, \country{USA}}}

\abstract{Artificial intelligence is beginning to reduce the manual effort in the CAD-to-mesh pipeline. Written for meshing and geometry practitioners with limited AI background, this survey organizes recent work by workflow step. We cover part classification and segmentation, mesh quality prediction, and defeaturing. We review AI guidance for unstructured meshing, block-structured meshing in 2D and 3D, and volumetric parameterization, including reconstruction from implicit or sampled geometry. We also discuss parallel mesh generation and scripting automation via reinforcement learning and large language models. Across these topics, AI complements established geometry and meshing algorithms rather than replacing them. We conclude with practical lessons and open challenges in data, benchmarks, and trustworthy integration.}

\begin{document}
\maketitle

\maketitle

\footnotetext[1]{Sandia National Laboratories is a multimission laboratory managed and operated by National Technology \& Engineering Solutions of Sandia, LLC, a wholly owned subsidiary of Honeywell International Inc., for the U.S. Department of Energy's National Nuclear Security Administration under contract DE-NA0003525.}

\section{Introduction}

High-fidelity simulation is now a cornerstone of engineering and scientific workflows. Yet, getting from a 3D model to a simulation-ready mesh remains one of the most time-consuming and frustrating steps in the process. Engineers still spend countless hours identifying parts, simplifying geometry, setting meshing parameters, and checking mesh quality. Many of these decisions are repeated across similar models, and small differences in geometry or intent can trigger large changes in meshing behavior. Even with mature geometry kernels and meshing algorithms, the day-to-day workflow can feel more like careful craftsmanship than a reliable pipeline.

This is not because the underlying algorithms are weak. It is because meshing sits at the intersection of geometry, analysis intent, solver requirements, and practical constraints. Much of what makes an expert effective is not a single technique, but the ability to choose among many options, to anticipate failure modes, and to apply a consistent strategy across a family of parts. That expertise is valuable, but it is often captured informally: in tribal knowledge, in scattered scripts, and in decisions that are hard to reproduce when the team changes or the model evolves.

At the 2025 International Meshing Roundtable, a panel of researchers and practitioners, including several authors of this paper, discussed whether artificial intelligence could help. The most useful framing was not replacement. The opportunity is assistance. AI can learn from prior examples, flag likely issues earlier, and automate repetitive steps, while conventional geometry processing and meshing remain responsible for robustness and guarantees. In other words, AI can help scale expert practice without pretending that meshing is a solved problem that can be handled by a single black box.

This paper is written for readers who are comfortable with geometry and meshing, but who may have little or no background in modern AI or machine learning. The goal is to make the landscape approachable, to connect each AI idea to a familiar meshing pain point, and to be clear about where the technology is already useful and where it still falls short. Throughout, AI and machine learning refer to data-driven methods that extract patterns from previous models and workflows, then use those patterns to make predictions or recommend actions. The details of specific algorithms matter, but the key question is simpler: what part of the meshing workflow becomes faster, more consistent, or easier to automate when experience is captured in data?

To help navigate the space, Table~\ref{tab:ml_methods_taxonomy} provides a taxonomy of the main families of AI methods that appear in geometry processing and mesh generation, along with where they show up in this paper. The table is meant as a roadmap rather than a prerequisite. Readers can treat it as a guide to the kinds of problems each method is being used to solve.

\begin{table}[!tb]
\caption{Taxonomy of AI and machine learning methods used in geometry processing and mesh generation workflows.}
\label{tab:ml_methods_taxonomy}
\begingroup
\renewcommand{\arraystretch}{1.05}
\small
\sloppy
\setlength{\tabcolsep}{4pt}
\centering

\newcolumntype{Z}{>{\raggedright\arraybackslash\hspace{0pt}}X}

\begin{tabularx}{\columnwidth}{
  >{\raggedright\arraybackslash}p{2.1cm}
  >{\raggedright\arraybackslash}p{3.0cm}
  Z
  >{\raggedright\arraybackslash}p{1.2cm}
}

\toprule
\textbf{AI/ML Category} & \textbf{Description} & \textbf{Referenced Sections} & \textbf{Citations} \\
\midrule

\textbf{Classical Machine Learning} &
Rule-based statistical learning (SVMs, decision trees, ensembles, surrogates) \cite{Hastie2009ESL,Cortes1995SVM,Breiman2001} &
3D Part Classification; Mesh Quality Prediction; Defeaturing; Parallel Mesh Generation &
\cite{IpRegli2006,Jayanti2006,sklearn,OwenIMR2019,OwenOSTI2022,OwenIMR2023,ParryIMR2023,Parry2025,gamatie2019,ipek2005,garner2025numa,garner2025dist} \\
\hline

\textbf{Deep Learning - Geometric Models} &
Neural networks specialized for geometry (CNNs, GNNs, Transformers, hybrids) \cite{LeCun2015DeepLearning,Bronstein2017GeometricDL,Vaswani2017Attention} &
3D Part Segmentation; Mesh Quality Prediction (GNN baselines); Volumetric Parameterizations; Scripting Automation (graph encodings) &
\cite{lambourne2021brepnet,dupont2022cadops,lee2023brepgat,wu2024aagnet,zheng2025sfrgnn,zhang2024brepmfr,dai2025brepformer,uwimana2025segmentation,jones2023self,hanocka2019meshcnn,Pfaff2021MeshGraphNets,SpraveDrescher2021,Wang2022MeshQualityGNN,jayaraman2021uv,wang2023hybrid,zhang2024brep2seq,wu2020comprehensive} \\
\hline

\textbf{Generative \& Representation Learning} &
Generative shape representations (implicit SDFs, parametric splines, hybrids, diffusion models) \cite{Kingma2014VAE,Goodfellow2014GAN,Ho2020DDPM} &
Unstructured Meshing (geometry construction); Volumetric Parameterizations (DL-Polycube, DDPM-Polycube, neural integral mapping) &
\cite{park2019deepsdf,mescheder2019occupancy,gropp2020implicit,peng2020convolutional,mildenhall2021nerf,larios2021deep,novello2022exploring,dong2024neurcadrecon,dong2025neurcross,zhang2023neurogf,liu2020neural,sharma2020parsenet,williamson2025neural,maruani2024ponq,he2025sparseflex,liao2018deep,chen2022neural,sundararaman2024self,DeepShapeCode,yu2025dlpolycube,yu2025DDPM,Zhan2025} \\
\hline

\textbf{Reinforce-ment Learning (RL)} &
Agents learn multi-step geometry and meshing actions via trial-and-error \cite{SuttonBarto2018RL,Mnih2015DQN,Haarnoja2018SAC} &
Defeaturing (future RL); Unstructured Meshing (quad, Delaunay); Block Structured Meshing (2D, 3D); Volumetric Parameterizations (sequence optimization); Parallel Mesh Generation (AMR control); Scripting Automation (block decomposition, thin volume reduction, CAD scripting); Language-Driven Scripting (integration with RL) &
\cite{pan2023reinforcement,tong2023srl,Thacher2025,diprete2024reinforcement,haarnoja-2018-sac,Patel2021AutoHex,Zhang2025CADRLHex,Gatti2022RLGNN,Yang2023RLAMR,owen2024cad,zhang2025reinforcement} \\
\hline

\textbf{Large Language Models (LLMs)} &
Text-to-code synthesis, multimodal assistants, and agentic workflows (LLM+tools, LLM+RL) \cite{Vaswani2017Attention,Brown2020GPT3,Ouyang2022InstructGPT}  &
Language-Driven Scripting; Scripting Automation (integration with geometric encoders and RL) &
\cite{zhao2023survey,R1,R2,R13,R14,R18,R26,N2,R9,R10,R12,R23,R25,fang2025selfevolving,Menon2026MultiAgentISC} \\
\bottomrule
\end{tabularx}

\endgroup
\vspace{-0.6em}
\end{table}

The range of applications is broad, but the motivation is consistent. Some methods help a workflow ``understand'' geometry in ways that support downstream decisions, such as recognizing parts, identifying functional regions, or separating features from context. Other methods help predict outcomes, such as whether a set of meshing choices is likely to produce poor elements or whether a local region is likely to cause failure. Still others are aimed at automation, where the goal is to turn an expert sequence of steps into something that can be repeated reliably, either as a learned policy or as generated scripts that drive existing CAD and meshing tools.

The remainder of the paper follows the workflow from geometric organization to mesh generation and automation. We first discuss part classification and segmentation, because these tasks often determine what information is available to drive meshing choices. We then survey mesh quality prediction and defeaturing, where AI can reduce costly iterations by warning about problems and suggesting targeted changes. Next, we cover unstructured and block-structured meshing, including recent work that uses learning to guide meshing decisions or to construct intermediate representations that enable structured discretizations. We also discuss volumetric parameterization methods, which sit at a key junction between geometric mappings and structured mesh generation. Finally, we address parallel mesh generation and automated scripting, including reinforcement learning and large language models that aim to make workflow automation more accessible to analysts.

Many of these efforts are still in early stages, but the momentum is clear. AI is beginning to reshape not only how meshes are generated, but also how geometry is prepared, validated, and integrated into simulation pipelines. At the same time, practical adoption highlights challenges that appear repeatedly across the literature: the need for curated datasets and shareable benchmarks, reliable representations of CAD geometry that preserve intent, and evaluation criteria that connect algorithmic success to simulation outcomes. This review highlights where AI is already making a difference, and it outlines what must improve for these tools to become dependable partners in everyday meshing practice.

\section{3D Part Classification}
%


Complex assemblies frequently include many repeated mechanisms such as bolts, screws, springs, and bearings. Analysts often spend substantial time identifying these parts and transforming them to prepare for analysis; for example, bolted connections may require specific geometric simplifications, specialized meshing, and boundary condition assignment. At the scale of hundreds of bolts per assembly, this manual process becomes tedious and error prone, which motivates automated 3D part classification as a front end to idealization and meshing.

Early work approached part classification with manual tagging and rule-based feature recognition. These systems encode expert heuristics about geometry or manufacturing features and can work for narrow classes, but they are brittle across diverse CAD sources, modeling conventions, and tolerances, and they scale poorly as category counts grow \cite{IpRegli2006,Jayanti2006}.

The authors of \cite{MCB2020} organize whole-part classification by data representation. In the multi-view image family, a CAD model is rendered from several viewpoints and the view set is classified; representative methods include MVCNN and RotationNet \cite{Su2015MVCNN,Kanezaki2018RotationNet}. In the volumetric family, tessellated solids are voxelized and processed with 3D CNNs; common baselines include VoxNet and VRN \cite{Maturana2015VoxNet,Brock2016VRN}. In the point family, methods learn directly from sampled points and often excel when local geometric detail matters; examples include PointNet++, PointCNN, and SpiderCNN \cite{Qi2017PointNetPlusPlus,Li2018PointCNN,Xu2018SpiderCNN}. These representation choices benefit from mature vision toolchains and large public datasets and they perform well for broad shape recognition, yet they may discard exact boundary information and can be sensitive to viewpoint, grid resolution, or sampling density \cite{MCB2020}. For simulation-driven workflows, these limitations motivate methods that keep the CAD boundary model intact.

CAD-native approaches preserve the boundary representation and map cleanly to idealization, meshing, and boundary-condition setup. In addition to supervised labeling, B-rep structure can also be used to build relational understanding at the assembly scale without committing to a fixed label set, which complements intent-centric CAD-CAE integration efforts based on explicit knowledge representations \cite{boussuge2019ontology}. Boussuge et al.\ use tensor factorisation over geometric and topological relations to support similarity-based retrieval and hierarchical grouping of entities and components in CAD assemblies \cite{boussuge2022tensorfactorisation}. While this is not semantic part classification, it can reduce analyst effort by organizing large assemblies into consistent groups that can later be mapped to task-specific categories when needed.

A practical CAD-native path computes tabular attributes from curves, surfaces, and volumes provided by a modeling kernel \cite{ACIS} and then trains a classifier such as a Random Forest \cite{Breiman2001} using scikit-learn \cite{sklearn}. Typical signals include counts by analytic face type and summary ratios over area, curvature, and edge or vertex statistics \cite{OwenIMR2023}. This design aligns with the cues analysts use, yields readable importance scores \cite{Breiman2001}, supports incremental updates with user-labeled examples \cite{OwenIMR2023}, and delivers strong accuracy with thousands of parts at assembly-scale inference speed \cite{OwenIMR2023,OwenOSTI2022}.

Formally, part classification can be cast as a supervised learning problem. Given a training set $\mathcal{D} = \{ (\mathbf{x}_i, \mathbf{y}_i) \}_{i=1}^{n}$, with feature vectors $\mathbf{x}_i$ that characterize each part and labels $\mathbf{y}_i$ that define the category, the objective is to learn a function $f$ such that $\hat{\mathbf{y}} = f(\mathbf{x})$. Early work used handcrafted descriptors with SVMs and established engineering benchmarks \cite{IpRegli2006,Jayanti2006}. Deep learning approaches for CAD classification followed \cite{Qin2014}. Recent studies also explore whole-part classification from B-rep-derived graphs that retain exact boundary semantics \cite{Roj2022CADGraphClassification,Mandelli2022CADGNN}. Labels are commonly drawn from single-part repositories and curated corpora such as GrabCAD, ABC, and MCB \cite{GrabCAD,Koch2019,MCB2020}. Comparative results show that tree ensembles remain competitive baselines on tabular B-rep features, offering fast and interpretable predictions suitable for deployment \cite{OwenIMR2023,OwenOSTI2022}.

Methods and implementation have matured into deployable tools. The Cubit Geometry and Meshing Toolkit integrates part classification and reduction features to drive predefined idealization recipes for meshing and boundary condition setup for fasteners \cite{CubitUserDoc1706_2025,OwenIMR2023,OwenIMR2019}. In a typical workflow, a complex assembly is first classified, the user reviews the resulting category assignments in the GUI, and then selects one or more categories (for example, bolts) to apply an appropriate recipe. These recipes encapsulate common actions for simulation preparation, including geometric transforms, meshing, decomposition, and boundary condition application. In Cubit, users can also retrain in situ by selecting new parts for categorization, adding new categories when needed, and persisting the resulting training data for future sessions; a small set of common categories is shipped by default and can be expanded as users contribute examples \cite{OwenIMR2023}.

Building on this pattern, other commercial preprocessors are adding related capabilities. Altair HyperMesh offers \emph{AI-driven shape recognition (ShapeAI)} that identifies and groups recurring parts such as fasteners \cite{AltairHyperMesh,AltairShapeAI}, and Siemens Simcenter NX incorporates shape recognition and similarity search to enable automated grouping and selection of geometrically related parts \cite{SimcenterAIWorkflow,SiemensNXShapeRecognition}. Adoption remains early, which underscores both the promise of these approaches and the still-limited extent of deployment across the broader simulation tool ecosystem.

Looking forward, promising directions include expanding class taxonomies beyond standard hardware, improving cross-CAD robustness, leveraging B-rep-native GNNs \cite{BRepNet,Colligan2022,Lee2023}, and coupling classification with downstream steps such as defeaturing and mesh-quality prediction to close the loop in data-driven model preparation.

\sectionauthor{S. Owen}

\section{3D Part Segmentation}

%

Segmentation of CAD models assigns semantic labels to regions of a part, most often at the face level. In contrast to part classification, which assigns a single label to a whole component (for example, ``bolt'' or ``flange''), segmentation identifies functional and geometric regions such as holes, fillets, chamfers, ribs, bosses, and contact or load-bearing surfaces. This finer structure is central to simulation workflows because it enables region-aware meshing, selective defeaturing, and reliable application of boundary conditions and material models. In practice, segmentation is also closely related to CAD feature recognition, especially when labels correspond to modeling operations or machining features.

Classical segmentation systems rely on rule-based feature detection using curvature, convexity, adjacency, and other geometric heuristics. These approaches can be effective within a narrow modeling style, but they tend to be brittle when applied across different CAD kernels, tolerance conventions, and feature interactions. Manual segmentation remains accurate, but it does not scale to the volume and iteration rate of modern engineering pipelines. Learning-based methods address these limitations by training models to recognize repeatable geometric and topological patterns directly from B-rep structure, and by producing labels with consistent semantics that can drive downstream automation.

Segmentation benchmarks differ not only in size, but also in what the labels represent. A useful split is between \textit{design-oriented} datasets, where labels reflect modeling operations and sometimes construction steps, and \textit{manufacturing-oriented} datasets, where labels follow machining features and other process-relevant attributes.

Design-oriented datasets align naturally with design intent. The Fusion 360 Gallery Segmentation dataset established a standard evaluation setting for face-level labels tied to common construction operations \cite{lambourne2021brepnet}. These operation-driven labels provide structured supervision that is meaningful for downstream automation. CC3D-Ops extends this style by adding construction step annotations that group faces created by the same operation, which supports sequence-aware reasoning \cite{dupont2022cadops}.

Manufacturing-oriented datasets align with process planning, and they also translate naturally to simulation preparation because machining features often define regions that require controlled mesh sizing, contact treatment, or localized refinement. MFCAD and MFCAD++ provide face-level machining feature labels, with MFCAD++ increasing geometric complexity and including more challenging feature interactions \cite{cao2020graph,colligan2022hierarchical}. More recent datasets add richer supervision, including instance grouping and bottom-surface identification, which makes the output closer to the structured selections engineers perform in practice. MFInstSeg is a representative example that supports semantic segmentation, instance segmentation, and bottom-surface labeling \cite{wu2024aagnet}.

Most modern CAD segmentation pipelines can be grouped by how they represent B-rep structure and how they propagate information across it. The common thread is to combine local geometric cues, such as surface type and convexity, with broader context implied by topology and feature interactions.

A widely cited baseline for direct B-rep learning is BRepNet \cite{lambourne2021brepnet}. Its key idea is to use directed coedges to define neighborhoods for topology-aware convolutions, allowing the network to update features on faces, edges, and coedges while respecting the B-rep structure. This topology-native design has been influential because it provides a clean way to combine geometry and connectivity without flattening the model into an unstructured point representation. Several extensions build on this foundation. CADOps-Net \cite{dupont2022cadops} jointly predicts operation types and construction steps, which helps connect segmentation outputs to design history. In machining-oriented settings, face labels can be grouped into coherent feature instances using connected component analysis after per-face classification \cite{cha2023machining}. For deployment constraints, pruning and transfer learning have been used to reduce computational cost and improve adaptation to new data distributions while retaining much of the baseline performance \cite{gkrispanis2025towards,gkrispanis2024enhancing}.

Graph-based approaches represent a B-rep as an attributed adjacency graph whose nodes correspond to faces (and sometimes edges), and whose edges encode topological relations. This representation makes it natural to use graph neural networks (GNNs) that propagate information across neighbors while learning which relationships matter most. BRepGAT \cite{lee2023brepgat} uses attention-based message passing to weight influential neighbors, which can help when local evidence is ambiguous. AAGNet \cite{wu2024aagnet} extends this direction with multi-task learning, predicting semantic labels alongside instance groupings and bottom-surface identification. These multi-task outputs align closely with what engineers need for downstream selection and automation, because they recover both feature type and feature extent in a form that can be consumed by rule-based or learned workflows. Domain shift remains a practical obstacle, since models trained on curated benchmarks can degrade on industrial parts. Approaches such as SFRGNN-DA \cite{zheng2025sfrgnn} target cross-domain robustness, while self-supervised and few-shot strategies aim to reduce dependence on dense manual labels when adapting to new feature libraries or conventions \cite{jones2023self}.

Transformer-based models have also been introduced to capture broader context across a part, which is helpful for interactive features, repeated patterns, and cases where local geometry is not sufficient to determine the correct label. BrepMFR~\cite{zhang2024brepmfr} combines transformer-style attention with graph structure and uses domain adaptation to improve transfer from synthetic to real CAD models. BRepFormer~\cite{dai2025brepformer} similarly propagates information through transformer layers while incorporating topological cues, and it has demonstrated strong performance on multi-task benchmarks such as MFInstSeg~\cite{wu2024aagnet}. In parallel, hybrid and multi-modal models combine complementary representations to balance robustness and expressiveness, for example by pairing UV-domain surface encodings with graph propagation~\cite{jayaraman2021uv}, mixing learned inference with rule-based decomposition for highly interactive features~\cite{wang2023hybrid}, or fusing B-rep and mesh features~\cite{uwimana2025segmentation,hanocka2019meshcnn}. Other directions address practical failure modes such as small-face underrepresentation through sampling-aware strategies and loss designs that improve fine-detail reliability~\cite{zhang2023extending}.

Most CAD segmentation systems remain research-stage, and broad commercial deployment is still limited. Even so, open-source baselines and better benchmarks are steadily improving reproducibility and lowering the barrier to integration. For simulation and meshing, the most impactful next steps are likely to be labels that are directly aligned with analysis decisions, such as identifying regions that require boundary-layer resolution, separating contact from non-contact surfaces, and predicting defeaturing candidates conditioned on target physics. Equally important is robust generalization across CAD kernels and modeling practices, since engineering pipelines depend on consistent behavior rather than best-case accuracy on a single benchmark. As these gaps close, segmentation can become a foundation for more automated CAD preparation, linking geometry understanding to defeaturing, mesh generation, and analysis setup.

\sectionauthor{N. Lei, X. Gu}

\section{Mesh Quality Prediction}
\label{sec:meshquality}

%

Anticipating mesh quality before meshing can save substantial time for complex CAD models because early identification of regions likely to yield poor elements prevents costly trial and error and focuses cleanup where it matters most. Rather than using meshing as a diagnostic that requires repeated cycles of mesh creation, inspection, control adjustments, and remeshing, an ML predictor can estimate local quality directly from CAD boundary-representation (B-rep) features prior to element generation. In practice, the most useful output is a ranked list or heat map of likely problem regions, which allows an analyst to target sizing, defeaturing, or partitioning actions early \cite{OwenIMR2019,OwenOSTI2022,OwenIMR2023}.

Mesh quality metrics are widely used in this setting because they are inexpensive, broadly applicable, and correlate with common failure modes such as inverted or highly distorted elements \cite{Knupp2000,Knupp2003,Verdict2007}. At the same time, mesh quality is only a proxy for what ultimately matters, namely the accuracy and reliability of simulation results. Learning simulation error or quantities of interest directly is often impractical because it requires solver outputs, boundary conditions, material models, and in many cases expensive reference solutions. A complementary line of work therefore predicts how model preparation choices affect downstream outcomes, such as estimating the impact of CAD defeaturing or simplification on heat transfer or CFD results \cite{Danglade2014,Danglade2015}. More broadly, surrogate modeling efforts learn to predict solution fields or derived quantities directly on meshes \cite{Pfaff2021MeshGraphNets}, but these methods are typically problem-specific and place stronger demands on training data and generalization.

As with part classification, mesh quality prediction is naturally posed as supervised learning. Given a dataset
$\mathcal{D} = \{ (\mathbf{x}_i, \mathbf{y}_i) \}_{i=1}^{n}$, the goal is to learn a function
$f$ such that $\hat{\mathbf{y}} = f(\mathbf{x})$.
The inputs $\mathbf{x}_i$ are constructed by assigning each B-rep entity (vertex, curve, and surface) a fixed-length feature vector that encodes local geometry and topology within a bounded neighborhood. Typical features include principal curvatures, dihedral angles, proximity to neighboring features, valence, and length and area ratios. This differs from part classification, where features are designed to characterize the part as a whole, often emphasizing global volume and overall shape. In mesh quality prediction, features are intentionally localized to the entity being evaluated and to the surrounding neighborhood that is most likely to influence the elements generated there.

The labels $\mathbf{y}_i$ are obtained by meshing a corpus of CAD parts under multiple target sizes and meshing settings, then assigning localized quality values near each entity, typically within one to two element edge lengths. Common scalar targets include minimum scaled Jacobian and minimum scaled inner radius, with definitions and implementations available in the standard literature and in libraries used by meshing tools \cite{Knupp2000,Knupp2003,Verdict2007}.

With inputs and labels defined, ensembles of decision trees (Random Forests) \cite{Breiman2001}, implemented in scikit-learn \cite{sklearn}, provide an effective baseline for these tabular B-rep features because they deliver strong accuracy, fast inference, and interpretable feature importances. Parry compares a random forest against a CNN for local quality prediction and reports better performance from the tree ensemble in this setting \cite{ParryIMR2023}, consistent with results reported in applied workflows \cite{OwenIMR2019,OwenOSTI2022,OwenIMR2023}. 

A secondary consideration, but an important one for tetrahedral meshing, is that quality outcomes can be inherently variable. Tetrahedral meshing is not space filling in the same way as structured approaches, and nondeterministic choices and small parameter changes can produce different local outcomes. This effect can be observed even on simple geometries: across repeated meshes of the same brick under random 3D rotations, the minimum scaled Jacobian for tets within two edge lengths of a surface varies by roughly 0.15. This variability motivates predicting a range, or the risk of failing a threshold, rather than a single deterministic score, and incorporating that uncertainty into how predictions are used \cite{Parry2026-LearningTetQuality}. Practically, this is supported by repeated meshing across settings, followed by aggregation into distributional targets such as quantiles or threshold probabilities \cite{Parry2026-LearningTetQuality}.

Related efforts outside CAD-to-mesh pipelines also cast mesh quality assessment as a learning problem, including element-level predictors and graph-based evaluations learned from mesh connectivity \cite{SpraveDrescher2021,Wang2022MeshQualityGNN}. These approaches are less central to CAD-driven pre-mesh decision making, but they reinforce the broader point that quality signals can be learned and can support automation.

The mesh quality predictor is implemented in the Cubit Geometry and Meshing Toolkit \cite{CubitUserDoc1706_2025}, where it flags low-quality regions prior to meshing and informs corrective actions that reduce iteration and accelerate time to analysis \cite{OwenIMR2019,OwenIMR2023}. Future work includes broader coverage of element types and quality metrics beyond tetrahedra, improved cross-mesher generalization, uncertainty-aware decision policies, and tighter coupling between mesh-quality risk predictors and downstream estimators of simulation impact, including learned surrogates and goal-oriented error indicators.

\sectionauthor{S. Owen}

\section{Defeaturing}
\label{sec:defeaturing}
%

CAD models intended for manufacturing often contain small features such as holes, fillets, embossed text, and fastener details. These features can be essential for fabrication, but they often have limited influence on the physics of interest and can make meshing expensive by forcing refinement and introducing poorly shaped elements. Defeaturing removes or simplifies such features so the mesh resolves what matters for analysis, without paying for unnecessary geometric detail.

Defeaturing in practice typically falls into three patterns. One option is to perform no defeaturing at all, allowing a conventional mesher to honor the full B-rep \cite{MeshGemsCADSurf}. This can produce far too many elements and unnecessarily small elements, increasing compute cost with little gain in accuracy. A second option is to handle defeaturing implicitly during meshing by using a feature-size tolerance or geometric envelope so the mesh effectively ignores small details \cite{Staten2024Morph,Hu2018TetWild}. This is usually fast and convenient, but it offers less control and can require iteration to confirm that important features were not suppressed. A third option is to explicitly edit the CAD model using geometry tools to remove or modify selected features. This offers the most control and often yields the best meshing outcome, but it is tedious and requires expertise. The latter two approaches are widely used in practice.

ML can reduce friction in tolerance-based meshing by predicting when it is likely to be safe, and it can accelerate explicit editing by identifying the entities most likely to create poor quality elements and recommending targeted CAD operations. In both cases, the goal is to reduce trial-and-error by estimating the expected meshing outcome, including the likely improvement, without requiring repeated meshing.

Both tolerant meshing and explicit CAD editing aim to reduce the defeaturing burden, but each comes with tradeoffs. Tolerant meshing can be an effective default in high-throughput workflows, and it is commonly available in commercial toolchains \cite{ANSYS2025}, but analysts still need safeguards to ensure that suppressed details do not change the physics of interest. Manual defeaturing tools such as CADfix \cite{CADfix2025} and SpaceClaim \cite{SpaceClaim2025} provide precise control, but they require significant analyst effort and deep experience with geometry operations and downstream meshing behavior. What is needed is a way to automate cleanup decisions that is both scalable and trustworthy, combining the speed of tolerant approaches with guidance and safeguards comparable to expert-driven editing.

Machine learning provides such an opportunity. By using mesh quality or simulation outcomes as supervision, ML models can learn to recognize problematic regions of geometry and evaluate corrective operations. Mesh quality is particularly attractive as a label because it is fast to compute, general across meshing tools, and provides a practical proxy for simulation performance. This allows thousands of training examples to be generated automatically, creating the data foundation needed for reliable prediction.  

Two recent studies illustrate different directions for ML-driven defeaturing. Shinde et al.~\cite{Shinde2024} supervised directly on simulation outcomes for sheet metal parts, labeling operations by their effect on quantities of interest in finite element analysis. This produces labels aligned with engineering objectives, but it requires many expensive simulation runs and limits the approach to specific physics. In contrast, Owen et al.~\cite{OwenIMR2019} proposed a more general framework driven by mesh quality prediction. Their method first estimates baseline quality for each B-rep entity, as described in Sec.~\ref{sec:meshquality}. Entities predicted to produce poor elements are flagged, and candidate defeaturing operations in Cubit \cite{CubitUserDoc1706_2025} such as \emph{remove surface}, \emph{composite surfaces}, \emph{blunt tangency} or \emph{collapse curve} are then evaluated by separate ML models that predict the post-operation quality. The system ranks both entities and operations, presenting analysts with prioritized cleanup options.  

Owen’s implementation~\cite{OwenIMR2019} relied on twelve feature models: nine corresponding to specific defeaturing operations and three for unmodified entities. Each model takes as input a fixed-length ``expert feature'' vector encoding local geometry and topology such as curvature, angles, adjacency, and area and length ratios. Training labels are mesh quality metrics including scaled Jacobian, scaled in-radius, and deviation. These labels are created by executing the CAD operation, meshing the modified geometry, and measuring quality near the affected entities. This process provides explicit evidence of how each operation impacts meshing so that at prediction time the model can recommend the best operation for a given local configuration. Training used the same ensemble decision tree method introduced in Sec.~\ref{sec:meshquality}, achieving low prediction error across the range of operations. The models were deployed both interactively and in automated greedy workflows, giving novice users a fast starting point while allowing experts to accept, reject, or refine suggested fixes.  

This capability has been released in Cubit \cite{CubitUserDoc1706_2025}  and is already in use, but it remains an active R\&D effort. Current work is expanding the feature models to cover more operations, improving generalization, and incorporating reinforcement learning. Whereas the current supervised approach predicts static outcomes for single operations, reinforcement learning would enable multi-step procedures that carry out full defeaturing end-to-end. Another direction is incorporating analyst preferences by mining journal files that capture real sequences of cleanup commands. These data could be used to adapt operation rankings to physics context. For example, structural dynamics analysts may favor different cleanup operations than those working in structural mechanics. Together, these efforts point toward defeaturing that is not only automated but also adaptive and informed by both geometry and human expertise.  

\sectionauthor{S. Owen}

\section{Geometry Construction from Implicit Data}
\label{sec:geometry-construction-implicit}

Defeaturing prepares CAD models for analysis by removing unnecessary complexity, but many simulation workflows do not begin with CAD. Medical imaging, reverse engineering, topology optimization, and scanned parts often provide geometry as voxels, implicit fields, or point samples rather than as parametric solids and surfaces authored in CAD systems. In these settings, the first meshing problem is not discretization, but \emph{domain construction}. The objective is to recover a well-conditioned geometric description that is watertight, faithful to important features, and stable under refinement so that downstream unstructured meshing and finite element analysis (FEA) do not inherit artifacts from the input representation. These non-CAD inputs are especially common in medical image-based modeling, where robust domain construction is a prerequisite for reliable analysis meshes~\cite{zhang2016geometric,zhang2013challenges}.

Classical pipelines convert implicit or discretized data into a boundary representation using methods such as Marching Cubes~\cite{lorensen1998marching}, Dual Contouring \cite{Zhang3DMeshing2005, ZhangHexMeshing2006}, and level-set extraction. These approaches are reliable workhorses, but their outputs can be sensitive to resolution, noise, and thresholding choices. For simulation, that sensitivity often appears as boundary roughness, inconsistent normals, and fragile sharp features, all of which complicate sizing, boundary-layer construction, and mesh quality control.

The methods reviewed here fall along three complementary axes: (i) continuous representations that decouple geometric fidelity from any single tessellation, (ii) differentiable reformulations of discrete extraction so reconstruction can be guided by downstream losses, and (iii) learned module replacement and end-to-end designs that retain trusted structure while improving robustness and adaptivity.

\textbf{Continuous representations.} Neural implicit models represent a shape as the zero level set of a learned field $f_\theta:\mathbb{R}^n \rightarrow \mathbb{R}$, with signed distance functions (SDFs), occupancy fields, and radiance-field variants being common choices~\cite{park2019deepsdf,mescheder2019occupancy,peng2020convolutional,mildenhall2021nerf}. Practical training strategies such as Eikonal regularization encourage well-behaved distance fields and improve conditioning~\cite{gropp2020implicit}. A key benefit for simulation pipelines is that continuous fields support differentiable geometric queries. Using automatic differentiation, normals and curvature-related quantities can be derived from $\nabla f_\theta$ and $\nabla^2 f_\theta$~\cite{goldman2005curvature,larios2021deep,novello2022exploring}. This supports feature-aware reconstruction that is more stable across discretizations, and it has been used in downstream workflows such as CAD reconstruction with improved sharp-feature retention and more robust cross-field generation for quad meshing~\cite{dong2024neurcadrecon,dong2025neurcross}. Continuous formulations can also encode distance structure on surfaces, for example neural geodesic fields that learn distances and paths, with decoupled variants improving scalability while maintaining accuracy in their reported settings~\cite{zhang2023neurogf,chencoupled}.

Neural \emph{parameterizations} provide a complementary continuous description by mapping low-dimensional coordinates to explicit surfaces such as NURBS, subdivision surfaces, or spline patches, giving stronger topological control and closer alignment with CAD workflows. ParSeNet predicts parametric control structures from raw 3D data to fit surfaces robustly~\cite{sharma2020parsenet}. Spherical Neural Surfaces provide a continuous formulation that stabilizes differential operators across irregular meshes and varying discretizations, improving the stability of spectral computations in that framework~\cite{williamson2025neural}. In addition, hybrid representations seek a practical compromise between flexibility and editability. PoNQ uses a learned quadric-error style representation to compress and reconstruct editable meshes efficiently~\cite{maruani2024ponq}, while SparseFlex leverages sparse neural fields to capture fine detail and thin structures under rendering or multi-view photometric supervision~\cite{he2025sparseflex}.

\textbf{Differentiable reformulations of extraction}. Even when geometry is represented continuously, most CAE pipelines ultimately require an explicit surface for boundary-conforming meshing. This motivates differentiable reformulations of classical extraction steps so that losses tied to geometry quality can influence reconstruction. Marching Cubes is non-differentiable because of discrete case tables and topology updates~\cite{lorensen1998marching}. Deep Marching Cubes relaxes thresholding and case selection into learnable surrogates, enabling supervision with geometric losses such as normal consistency while still producing a mesh output~\cite{liao2018deep}.

Building on this idea, related work explores alternative differentiable isosurfacing formulations that better align extraction with optimization and downstream objectives. Deep Marching Tetrahedra (DMTet) uses a tetrahedral discretization and differentiable marching tetrahedra to extract high-resolution triangle meshes from learned fields~\cite{shen2021dmtet}. FlexiCubes targets stability and mesh quality during gradient-based optimization by introducing a flexible isosurface extractor that improves robustness when extraction is embedded in learning or inverse-design loops~\cite{shen2023flexicubes}. Dual Contouring can likewise benefit from learned vertex placement and self-supervised geometric consistency, reducing dependence on ground-truth meshes while improving sharp-feature preservation~\cite{chen2022neural,sundararaman2024self}. These relaxations provide usable gradients, but practical systems must still control instability in local topology decisions to avoid frequent flips and inconsistent surfaces~\cite{liao2018deep,chen2022neural,sundararaman2024self}.

\textbf{Learned modules and end-to-end designs.} A third axis replaces brittle or hand-tuned components with learned modules, while still producing explicit geometric outputs and preserving the overall pipeline structure. Learned vertex placement in Dual Contouring is a representative example, where the neural component replaces heuristic optimization and improves sharp-feature preservation under realistic input variability~\cite{chen2022neural,sundararaman2024self}. Hybrid designs such as SparseFlex similarly couple a neural representation to multi-view photometric constraints, allowing reconstruction quality to be driven by dense supervision signals without requiring ground-truth meshes~\cite{he2025sparseflex}. For CAE, the appeal of these designs is pragmatic: they retain familiar stages (reconstruction, extraction, meshing) while reducing manual cleanup and improving robustness across acquisition modalities.

In a complementary direction, recent generative models produce explicit surface meshes directly. Meshtron~\cite{hao2024meshtron} demonstrates high-fidelity triangle mesh synthesis at scale, which is best viewed in this context as a geometry representation step. For simulation workflows, such outputs may still require conditioning, validation, or remeshing to meet analysis constraints.

From a technology readiness perspective, geometry construction from implicit data is comparatively mature in computer graphics, but its translation into routine CAE workflows remains at an early stage. The opportunity is clear: better-conditioned domains reduce manual cleanup, improve meshing robustness, and make it easier to connect reconstruction to simulation objectives. Near-term progress is likely to focus on topology control and feature-aware validation tailored to analysis needs, together with tighter integration into downstream sizing, defeaturing, and mesh-quality prediction.

\sectionauthor{N. Lei, X. Gu}

\section{Unstructured Mesh Generation}
\label{sec:unstructured-mesh-generation}

Once a domain is available, either from CAD or constructed from implicit data as described above, the workflow bottleneck shifts to unstructured mesh generation. Mesh robustness and element quality directly influence accuracy, efficiency, and convergence in FEA. Classical unstructured meshing algorithms are mature and widely deployed, but they rely on heuristic decisions that can be brittle across geometry classes and often require manual intervention to achieve reliable quality.

Delaunay-based methods, advancing-front techniques, and paving remain central for tetrahedral, triangular, and quadrilateral meshing. These methods combine proven geometric structure with a sequence of design choices, including point insertion, front updates, smoothing, and local refinement. In practice, it is these choices, rather than the high-level algorithm, that limit robustness. Small changes in geometry, sizing fields, or boundary complexity can degrade element quality or trigger topological failures, and the resulting fixes are often manual and time consuming.

The main opportunities for learning arise in the parts of unstructured meshing where heuristic choices control success or failure.  Heuristic rules are typically designed for common cases, whereas production geometries are diverse and increasingly originate from non-CAD sources. Learned policies can adapt to local geometric context while still operating within classical algorithmic frameworks, preserving structural guarantees while replacing brittle decision logic with data-driven components.

One representative line of work formalizes meshing as a sequential decision process. Advancing-front all-quad meshing can be posed as reinforcement learning, where a Soft Actor-Critic policy learns front updates that generalize across boundary conditions and sizing fields \cite{pan2023reinforcement}. Building on this idea, SRL-Assisted AFM combines supervised pretraining with reinforcement-learning fine-tuning to control singularities, improving cross-field alignment and local element quality on challenging planar domains \cite{tong2023srl}. In practice, these methods target robust quad layouts in multiply connected regions by making singularity placement an explicit part of the learned control policy, which helps maintain consistency of the front progression while preserving element quality near holes and other constraints.

Related work also applies learning to Delaunay-style triangular meshing, using graph-based policies to guide point placement and connectivity while preserving the underlying Delaunay structure~\cite{Thacher2025}. In parallel, several efforts embed learning into established pipelines rather than replacing them end-to-end. Neural subdivision variants retain Loop or Catmull--Clark combinatorics and learn residual vertex updates conditioned on local geometry~\cite{liu2020neural}, which can improve fairness and feature preservation during refinement. Hybrid pipelines also demonstrate how classical kernels can be paired with learned refinement while keeping trusted structure intact. Quadify is a representative example that integrates differentiable Catmull--Clark refinement into a pipeline that builds on field-aligned meshing foundations such as Instant Field-Aligned Mesh~\cite{fruhauf2024quadify,jakob2015instant}. Task-driven simplification provides another instructive pattern: MeshCNN learns edge importance so that decimation preserves structures relevant to downstream objectives~\cite{hanocka2019meshcnn}. In simulation settings, the analogous opportunity is learned refinement and coarsening guided by solution features and error indicators, with the learned component acting as an adaptive policy inside a verified meshing framework.

AI-enhanced unstructured meshing remains at a low technology readiness level. Most approaches are research prototypes, and practical adoption must address verification, robustness across geometry distributions, and compatibility with CAE requirements such as boundary layers, sizing constraints, and quality guarantees. The potential impact is significant because reducing manual intervention in meshing directly lowers analysis turnaround time and makes larger design spaces tractable. Promising directions include AI-guided point insertion and refinement in tetrahedral meshing, learned refinement and coarsening informed by error estimation, and tighter coupling between meshing decisions and simulation feedback. Over time, these advances can help move unstructured meshing from a heuristic bottleneck toward a more adaptive, objective-aware component that integrates smoothly with defeaturing, segmentation, and downstream FEA.

\sectionauthor{N. Lei, X. Gu}

\section{Structured Meshing}
%
Many advanced physics solvers, such as those for shock physics or structural dynamics, depend on meshes composed entirely of quadrilaterals (2D) or hexahedra (3D). Solver accuracy and stability can be highly sensitive to mesh structure, making all-hex and block-structured approaches especially valuable \cite{Benzley1995,Schneider2022}. The principle is straightforward: decompose a complex domain into four-sided or six-sided blocks and fill each with a structured grid, often using transfinite interpolation (TFI) \cite{Thompson1998}. A widely adopted variant, the ``pave-and-sweep'' strategy \cite{StatenCanannOwen1999,RocaSarrate2010}, generates a quadrilateral surface mesh and extrudes it through a third dimension to form hex elements.

Over decades, geometry- and topology-based methods such as mapping, submapping, and pave-and-sweep have provided the foundation of structured meshing. Production tools embody these strategies—Sandia’s Cubit \cite{CubitUserDoc1706_2025} integrates an interactive environment for decomposition, while industrial packages like \emph{Gridgen} \cite{Chawner1995} and \emph{GridPro} \cite{Ferlemann2010} emphasize smoothing and topology-first multiblock generation for CFD. Yet despite their maturity, these methods leave the hardest step—discovering a valid block decomposition—largely in the hands of expert analysts. Progress in 2D has been tangible, but robust, general-purpose automation in 3D has remained elusive.

Structured meshing also includes field-aligned quadrilateral layouts on surfaces, where the goal is to align elements with global shape and anisotropy rather than to define an explicit multiblock parameterization. Frame fields \cite{Ray2006NSymmetry} provide a compact way to encode these preferred directions, guiding quadrangulation toward coherent alignment with curvature and feature lines while concentrating unavoidable topological transitions into a small set of singularities. Recent learning-based work targets this field construction step directly. NeuFrameQ predicts neural frame fields and uses them to drive scalable anisotropic quadrangulation, emphasizing robustness across diverse shapes \cite{Liu2025NeuFrameQ}.

That humans can routinely identify good decompositions where algorithms fail points to what is missing: abstract reasoning, adaptive decision-making, and the ability to plan sequences of operations. Early attempts to inject AI into meshing, such as expert systems proposed by Lu, Gadh, and Tautges \cite{lu2001feature}, hinted at this direction but lacked the flexibility to generalize. Modern machine learning, and RL in particular, offers a more powerful framework. RL agents are explicitly designed to learn action sequences that optimize long-term goals, making them well aligned with the challenges of block decomposition, where the aim is to maximize completeness, mesh quality, and element conformity.

\subsection{2D Block-Structured Meshing}
While 2D block decomposition has direct applications, one of its roles today is as a proving ground for AI methods aimed at the more complex 3D case. DiPrete, Garimella, Cardona, and Ray \cite{diprete2024reinforcement} recently reframed the problem using RL. Their agent learned to apply sequences of axis-aligned cuts to orthogonal polygons, decomposing them into quadrilateral blocks. The framework was built on Soft Actor–Critic (SAC) \cite{haarnoja-2018-sac}, whose maximum-entropy objective promotes exploration, stabilizes training, and supports continuous state and action spaces.

The state representation captured local geometry around vertices in fixed-length vectors, while the reward function encouraged quad generation with uniform geometric properties (e.g., area, aspect ratio) and lightly penalized unproductive actions. On axis-aligned datasets restricted to $X$- and $Y$-direction cuts, the agent consistently achieved $100\%$ quadrilateral decompositions. Follow-on work expanded the action set with bisections and generalized the approach to arbitrary polygons \cite{RayRL2025}. For orthogonal shapes, the policy learned to favor edge-aligned cuts that produced rectangles; for general polygons, it generated mostly quadrilaterals but sometimes left residual triangles. Of course, these triangles could be decomposed into three quadrilateral blocks but those would be sub-optimal. These results show that RL can capture effective decomposition strategies, but fully generalized behavior remains out of reach.

The limitations also highlight a key distinction from standard RL benchmarks: the meshing environment is not static but varies dramatically across CAD models. In particular, the number and type of actions to apply are state-dependent and can vary significantly. Capturing such variability with fixed-length features is brittle and often leads to noisy training. Promising directions include richer, multi-scale state representations, larger and more diverse training corpora, and the use of imitation learning \cite{Argall2009SurveyLfD} to provide strong initial policies. In this way, 2D decomposition continues to serve not only as a useful capability in its own right but also as a critical stepping stone toward 3D block-structured and all-hex meshing, where the challenges—and the potential impact—are far greater.

An alternative to the approach taken in \cite{diprete2024reinforcement} is to leave the geometry domain unchanged and build a block structure that discretizes it. This can be achieved by considering the work of Naranayan et. al.~\cite{Narayanan2024}, who present a learning-based framework for improving the quality of unstructured triangular and quadrilateral meshes. Their approach learns to improve mesh quality via self-play RL with no prior heuristics.  They use an actor-only version of the Proximal Policy Optimization (PPO) algorithm (see~\cite{schulman2017}) without a critic (i.e., value function) network. They demonstrate their agent's ability to transform unstructured quadrilateral meshes—obtained by splitting triangles into four quads—into structured meshes that correspond to 2D block decomposition. Although the goal of~\cite{Narayanan2024} is not block decomposition, the applicability of their approach opens promising perspectives for pursuing this path.

Complementing these learning-based strategies, Bourmaud et~al.~\cite{PaulMCTS2025} frame 2D decomposition as \emph{Monte Carlo Tree Search} (MCTS): starting from the CAD model as root, the search expands candidate partition operations while balancing exploitation and exploration. The method often finds near-optimal quadrilateral decompositions, showing that high-quality layouts can emerge without end-to-end training. Extending MCTS to 3D is especially promising—partial block layouts alone could guide experts and scaffold downstream learning. More broadly, search-based methods complement RL when reward shaping and state design are brittle, and they anticipate the demands of 3D all-hex and block-structured meshing: long-horizon reasoning under sweepability, feature-preservation, and quality constraints, developed next.

\subsection{3D All-Hex and Block Structured}
All-hex meshing of complex CAD models begins with a sound geometric decomposition. Downstream schemes, including unstructured approaches (for example, pave and sweep), structured mappings (for example, map), and block-structured layouts, then generate conforming, high-quality hexahedra. The central difficulty is not meshing a single sweepable or mappable block, but discovering a \emph{valid} global decomposition and an executable sequence of operations that (i) respect sweepability constraints, (ii) preserve curvature and sharp features, and (iii) maintain element quality. Decades of practice have codified this workflow that provides robust predicates and operations, yet the pivotal decomposition step remains largely manual. This gap motivates AI methods that can capture expert strategies, reason over long action horizons, and adapt decisions to complex geometry.

Recent work has begun to push in this direction. Patel et al.\ \cite{Patel2021AutoHex} introduced Auto-Hex, which integrates geometric reasoning with RL. Starting from a boundary representation (B-Rep) and chordal-axis skeleton, the system identifies features such as T-junctions, sweep directions, and through-holes. An RL policy then proposes webcut actions conditioned on these cues, with feedback from the meshing environment guiding the agent to favor decompositions that yield hex or hex-dominant meshes. In related 2D work discussed earlier, DiPrete et al.\ \cite{diprete2024reinforcement} used CAD shape testbeds to study RL-driven cut sequencing and reward design, while Zhang et al.\ \cite{Zhang2025CADRLHex} frame 3D decomposition as a sequential decision problem and employ Soft Actor-Critic (SAC) to generate sweepable sub-volumes, showing promising decompositions on complex solids while highlighting open challenges in reward shaping and mesh quality.

In parallel, supervised learning has been explored as another route to automation. Quadros et~al.\ \cite{Quadros2025Bipartite,QuadrosIMR2026} reduce a general 3D CAD model (outer BRep “skin”) to a compact 1D bipartite, heterogeneous graph derived from a 2D chordal axis skeleton (CAT), then apply a heterogeneous GNN to predict decomposition operations. Each node (surface or curve) carries geometric descriptors (e.g., area, length, centroid, thickness), topological cues (valency, interior loops), and meshing indicators (principal sweep directions), and the pipeline maintains bidirectional maps between BRep, CAT, and graph so predictions can be emitted as executable WebCut strings. Because labeled decompositions are costly, training is performed on randomized, synthetic CAD with multi-action labels; on held-out lattices the approach attains $98.72\%$ per action accuracy. Finally, they demonstrate effective performance on out-of-distribution geometries, achieving $85\%$ accuracy for WebCut predictions on complex industrial 3D CAD models.

Beyond aggregate accuracy, a key capability of this representation is that predictions can be traced back to local geometric and topological evidence in the graph. By propagating information over multi-hop neighborhoods, the GNN can learn recurring motifs that correlate with useful decomposition operations, such as cuts that expose sweep directions or isolate regions that are difficult to mesh. In practice, this shifts the workflow from manual exploration of many possible operations to a short list of plausible, high-impact candidates that can be applied directly through the mapped WebCut actions. Ongoing work extends this supervised framework to targeted CAD families and couples it with reinforcement learning, in the spirit of Auto-Hex agents, to learn action sequencing on general geometries without relying exclusively on labeled data.

Successful all-hex meshing is rarely a linear pipeline. Instead, it behaves like a tightly coupled loop in which decisions made early are repeatedly revisited as later checks expose new constraints. Practitioners typically cycle through physics-aware cleanup and defeaturing, specification of parting or symmetry, geometry decomposition via WebCuts, imprint and merge, sweepability checks and interval assignment, mesh generation, and quality assessment with localized repair, before attribution and model handoff. In practice, downstream checks frequently force returns to earlier stages. For example, imprint operations can introduce slender faces that require revised cuts, residual subvolumes may fail sweep checks and trigger alternative decompositions, and localized quality failures often motivate targeted changes to the cut plan rather than global retuning.

Taken together, these efforts show clear progress but also uneven maturity. In 2D, RL agents reliably decompose orthogonal polygons into quadrilaterals and show promising, if less consistent, performance on general shapes \cite{diprete2024reinforcement}. In 3D, methods such as Auto-Hex \cite{Patel2021AutoHex}, SAC-based sequential decomposition \cite{Zhang2025CADRLHex}, and supervised GNN approaches \cite{Quadros2025Bipartite} demonstrate feasibility on curated models and strong results on synthetic datasets, but remain at an early research stage and are not yet ready for routine use on complex industrial geometries.

The most promising path forward lies in hybridization and scale. Supervised models can provide strong priors for plausible decompositions, while RL agents can optimize their ordering and long-term consequences. Richer geometric encodings, larger and more diverse CAD corpora, and tighter coupling to production meshing environments will all be essential. Even partial automation—where algorithms propose cuts that analysts validate—could reduce manual effort from weeks to hours, broaden access to structured meshing, and improve solver robustness through higher-quality meshes. Looking further ahead, the integration of autonomous decomposition agents directly into meshing tools points toward the long-sought goal of true “push-button” all-hex meshing for classes of geometries that remain out of reach today.

\sectionauthor{N. Ray, R. Garimella, R. Quadros, F. Ledoux}

\section{Volumetric Parameterizations}

%



Volumetric parameterization maps a solid 3D domain to a simpler parametric volume, often a unit cube, a set of cuboids, or a polycube. In simulation workflows, this mapping is not an abstract convenience. It is a bridge from geometric shape to analysis-ready structure. For isogeometric analysis (IGA), many spline constructions rely on tensor-product bases, so a hexahedral control mesh and an analysis-suitable volumetric parameterization are closely linked \cite{cottrell_isogeometric_2009}. For complex geometries, producing a bijective map with low distortion remains difficult, and that difficulty directly impacts numerical stability and accuracy in downstream analysis \cite{zhang2016geometric}.

Unlike block-structured meshing, which decomposes a geometry into discrete blocks to be meshed locally, volumetric parameterization seeks a continuous global map. When it succeeds, it enables two outcomes that are hard to obtain by purely local decisions: (i) all-hex meshes that act as control nets for spline construction, and (ii) tensor-product spline volumes whose structure aligns naturally with IGA discretizations \cite{cottrell_isogeometric_2009}. When it fails, the failure modes are often decisive for simulation, including foldovers (negative Jacobians), excessive distortion, or singular structures that propagate into poor element quality and ill-conditioned systems.

A common theme in classical approaches is to enforce alignment to geometry and boundaries while also controlling injectivity and distortion. CubeCover \cite{Nieser2011CubeCover} uses volumetric frame fields together with integer-grid maps to construct boundary-aligned parameterizations, illustrating how discrete constraints and continuous maps must be balanced to avoid inversions. Another widely used route is polycube-based parameterization, which deforms the input into a polycube domain and then pulls back a regular grid to obtain a structured hex layout \cite{Gregson2011Polycube,Fang2016ClosedFormPolycube}. Recent work improves robustness and consistency through intrinsic formulations \cite{Mandad2022IntrinsicPolycubes} and quantization strategies \cite{Brueckler2022Quantization}, aiming to better control alignment while reducing artifacts that degrade hex quality. Despite progress, these pipelines still face recurring bottlenecks on engineering models: layout decisions can be sensitive, optimization can be expensive, and user guidance is often needed to avoid poor decompositions and distorted regions \cite{zhang2013challenges}.

In IGA and geometry processing, parameterization is often framed as constructing spline-ready volumes rather than only extracting elements. Early approaches convert unstructured quad and hex meshes into T-spline surfaces and volumes \cite{wenyan2010a,Wenyan2011a}. Harmonic and Laplacian-based formulations with Dirichlet boundary conditions have also been used to define mappings \cite{Xia2010}, supporting solid T-splines for genus-0 \cite{zhang_solid_2012,Wenyan2013c} and higher-genus domains \cite{LeiLiu2012a}. Polycube decomposition has been particularly influential for IGA because it provides a practical route to multi-cuboid parameter domains that can be parameterized by B-splines \cite{LZH2014,Liu2015,HZ2016CMAME,HZL2017}. Alongside this, a range of spline technologies have been incorporated into analysis pipelines, including weighted T-splines \cite{LeiLiu2015a}, truncated T-splines \cite{ref:wei16a}, and truncated hierarchical B-splines \cite{wei17a,wei2018blended}. These ideas have also reached commercial and production settings \cite{Lai2016,Lai2017,yu2022hexgen}, which underscores both the importance of the problem and the need for robust automation.

Across these families, the recurring challenges are consistent: computational cost, sensitivity to initialization and layout decisions, user dependence, and difficulty guaranteeing quality. For many geometries that arise in engineering, ensuring bijectivity while keeping distortion low is the central obstacle. When the mapping introduces poor elements, downstream optimization is often limited, and the resulting representations may be unsuitable for stable simulation.

AI is a good fit for volumetric parameterization for the same reason it is a good fit for earlier steps in the pipeline covered in this review. Practical parameterization depends on discrete, high-level decisions that are difficult to encode with fixed heuristics, including layout selection, surface labeling, and choosing boundary correspondences that balance alignment with distortion. Learning-based methods can propose these decisions quickly, while geometry-aware optimization can enforce smoothness and validity constraints.

DeepShape \cite{DeepShapeCode} voxelizes the input and uses convolutional networks to establish correspondence with voxelized polycubes, reducing reliance on hand-tuned heuristics in the mapping process. Evocube \cite{Dumery2022} explores evolutionary search to label faces and guide polycube construction, illustrating a complementary direction where AI explores a large discrete decision space that is otherwise difficult to navigate reliably. More recently, DL-Polycube \cite{yu2025dlpolycube} predicts a polycube structure and uses it to guide surface segmentation, then combines octree subdivision, parametric mapping, and quality improvement to produce an all-hex mesh and an IGA-ready volumetric spline representation. This end-to-end pipeline is representative of how learning can reduce manual effort in polycube construction while keeping the downstream mapping and improvement stages grounded in classical geometry processing.

A second AI direction is to represent the parameterization using neural functions, which can serve as flexible approximators for complex mappings while incorporating geometric constraints. Penalty-based formulations remain important for controlling distortion and discouraging foldovers \cite{Ji2022}, and neural models can be used either to provide strong initializations or to learn components of the mapping that are expensive to derive from scratch. The integral parameterization approach in \cite{Zhan2025} uses neural networks to represent an inverse parameterization while enforcing boundary correspondence and interior mapping, directly targeting the bijectivity and distortion tradeoffs that limit classical pipelines. Diffusion-based polycube generation has also been explored as a generative route to robust layouts, with DDPM-Polycube \cite{yu2025DDPM} modeling polycube creation as a denoising process for producing high-quality hex meshes and volumetric splines.

From a readiness standpoint, the most reliable near-term pattern is hybrid: learning proposes layouts, segmentations, and initial correspondences quickly, then geometry-aware optimization enforces smoothness and validity. The key open gap for broad adoption in simulation is not only speed, but predictable behavior across diverse CAD topologies and feature scales \cite{zhang2013challenges}. This places a premium on verification strategies for injectivity, robust handling of thin regions and complex topology, and benchmarking on engineering-relevant model collections. As these pieces mature, AI-empowered volumetric parameterization is positioned to reduce the cost of building analysis-suitable spline volumes and all-hex control meshes, expanding the practical reach of IGA in applications such as biomedicine, materials, and engineering design \cite{cottrell_isogeometric_2009}.

\sectionauthor{Y. Zhang}

\section{Parallel Mesh Generation}

%

As simulation fidelity and geometric complexity rise, large-scale FEA increasingly hinges on \textbf{parallel mesh generation} to sustain simulations with hundreds of millions of elements~\cite{chrisochoides2006}. Misconfigured runtimes can waste substantial wall time at this scale, so choosing the right parameters matters as much as the mesher itself. Both data and computation must be distributed over many processors in an HPC environment. Modern parallel meshers are indispensable, but they expose numerous runtime controls such as block size, work-partitioning strategy, \textbf{NUMA affinity}, and communication thresholds, and these choices have first-order effects on scalability and wall time. Selecting an optimal parameter vector $p$ is a nonlinear and architecture-sensitive problem that has traditionally depended on expert heuristics or costly trial and error.

Analytic performance models exist, but they typically rest on simplifying assumptions and generalize poorly across architectures and problem families~\cite{barker2005practical}. Empirical tuning is more faithful but costly: each evaluation consumes a full HPC submission, and queue delays often dominate turnaround time. Collecting even a few dozen labeled runs can take weeks or even several months in production settings~\cite{angelopoulos2018}. These constraints impede routine optimization when rapid results are needed.

Artificial intelligence offers a complementary path. In particular, supervised learning can train \textbf{surrogate models} that map parameter vectors to execution time~\cite{gamatie2019, ipek2005}. Training shifts cost up front. Once learned, the surrogate supports fast what-if exploration over unseen configurations without consuming additional HPC cycles, which enables wide synthetic searches of the parameter space.

As one illustrative case study, compact feedforward neural networks were trained as surrogate models to predict the execution time of a general-purpose parallel mesher from a parameter vector $p$ using small real-world datasets~\cite{garner2025scalability}. In that study, experiments were reported for two execution environments: a shared-memory cc-NUMA system with 172 runs~\cite{garner2025numa}, and a distributed-memory setting with 68 runs~\cite{garner2025dist}. The distributed-memory results drew on representative node configurations including (i) dual-socket AMD EPYC 7763 nodes @ $2.45$\,GHz ($64$ cores per socket, $128$ total slots) with $256$\,GB memory, and Intel Xeon nodes such as (ii) dual-socket Xeon Gold 6148 @ $2.40$\,GHz ($20$ cores per socket, $40$ total slots) with $384$\,GB memory, and (iii) dual-socket Xeon E5-2698 v3 @ $2.30$\,GHz ($16$ cores per socket, $32$ total slots) with $128$\,GB memory. Reported results indicated that hierarchical load balancing was particularly beneficial on cc-NUMA platforms, consistent with the importance of locality for memory-bound stages~\cite{garner2025numa, garner2025dist}.

In the same case study, the trained models scored $100{,}000$ synthetic configurations in under $0.5$\,s per case, turning weeks of batch exploration into minutes. On held-out tests, the models reached a Mean Absolute Percentage Error (\textbf{MAPE}) of $2.13\%$ on the shared-memory system and $5.68\%$ on the distributed-memory system~\cite{tsolakis2024, garner2025dist}. This accuracy was sufficient to select settings that reduced total execution time. In practical terms, the primary bottleneck in HPC workflows is often cumulative queue wait time, and the study estimated an \textbf{about $98\%$ reduction} in total time-to-solution relative to a conservative manual search that would require tens or hundreds of queued jobs.

The core challenge in optimizing performance lies in the complexity of modern meshing codes. The workflow is decomposed into multiple, tightly coupled, parallel stages, where the optimal parameter choice for one stage is often dependent on the choices made for others~\cite{chrisochoides2006}.

\textbf{Domain Decomposition:} The initial geometric domain is partitioned into smaller subdomains for concurrent processing by different processors~\cite{chrisochoides2006}. The decomposition strategy significantly influences communication patterns and load balancing, where a poor choice can cause substantial overhead~\cite{barker2005practical, thomadakis}. In addition to methods like recursive coordinate bisection, geometry-based techniques such as PQR (a sorting-based method utilizing a curvilinear coordinate system that binds to the boundary for partitioning elements) were also used in specific test cases~\cite{chrisochoides1994, garner2025dist}. The performance impact is determined by the surface-to-volume ratio of the subdomains.

\textbf{Parallel Meshing and Refinement:} Each subdomain is meshed simultaneously using core methods like Delaunay triangulation~\cite{Foteinos_JPDC_2014}. Adaptive isotropic mesh generation employs parameters such as $\delta$ (determines element size)~\cite{Foteinos_JPDC_2014} along with $nthreads$ and an adaptation heuristic such as $adapt-const$~\cite{tsolakis2024, garner2025dist}. Parameters control criteria such as point insertion and topological modifications (e.g., local reconnection flips)~\cite{tsolakis2024}.

\textbf{Dynamic Load Balancing:} A runtime system dynamically manages the distribution of work to mitigate computational load imbalance~\cite{thomadakis}. The implemented hierarchical load balancing model specifically targets cc-NUMA architectures. It prioritizes assigning workloads to threads that are pinned to CPU cores within the shortest NUMA node distance, effectively reducing costly remote memory accesses~\cite{garner2025numa, garner2025dist}.

The surrogate model uses a feature vector $\mathbf{x}$ comprised of algorithmic, quality, and parallel parameters to predict the execution time. This parameter space is large and complex, as summarized in Table~\ref{tab:parameter_summary}, drawing from the general problem structure.

\begin{table}[htbp]
\footnotesize
\centering
\caption{Examples of Key Parameter Categories Influencing Parallel Mesh Generation Performance}
\label{tab:parameter_summary}
\setlength{\tabcolsep}{3pt} 
\begin{tabularx}{\linewidth}{@{}l>{\raggedright\arraybackslash}X>{\raggedright\arraybackslash}X@{}}
\toprule
\textbf{Category} & \textbf{Examples of Parameters} & \textbf{Primary Impact on Performance} \\
\midrule
\textbf{Algorithmic} & Meshing technique (Delaunay) & Controls efficiency and quality guarantees. \\
& Execution model (eg. speculative~\cite{Foteinos_JPDC_2014}) & Affects conflict resolution overhead (e.g., rollbacks). \\
\midrule
\textbf{Quality \& Fidelity} & Element quality criteria & Determines stopping criteria. \\
& Sizing control method & Dictates final mesh density and complexity. \\
& Surface fidelity control~\cite{Drakopoulos2024} & Deviation tolerance from input geometry. \\
\midrule
\textbf{Parallelism} & Domain decomposition method & Influences communication patterns. \\
(Data/Task) & \textbf{Load balancing scheme} & Minimizes processor idle time. \\
& Task granularity~\cite{Tsolakis2022} & Ratio of computation to communication. \\
\midrule
\textbf{System/Runtime} & Inter-processor communication & Time spent exchanging boundary data. \\
& Load imbalance / quality & Uneven work or poor initial partitioning. \\
& Number/type of processors/cores~\cite{Thomadakis2022} & Total parallel resource count. \\
\bottomrule
\end{tabularx}
\end{table}

Beyond surrogates, recent work uses learned modules to dynamically decide about: (1) load balancing, where trained classifiers select domain-decomposition settings and reduce time-to-solution compared to static heuristics for production workflows like \texttt{snappyHexMesh}~\cite{Gangopadhyay2025ParCFD}; (2) decomposition and partitioning, where reinforcement learning (RL) with graph neural networks (GNNs) learns partitioners that compete with multilevel methods while providing a trainable interface to work distribution and communication footprint~\cite{Gatti2022RLGNN}; and (3) adaptation, where adaptive mesh refinement (AMR) is posed as a sequential decision problem, and learned policies make locality-aware refinement choices that scale across partitions and generalize across PDEs and mesh sizes~\cite{Yang2023RLAMR}.

Together, these results suggest a practical division of labor: surrogates accelerate search over interdependent runtime parameters, while learned controllers directly optimize decomposition, load balancing, and adaptive refinement. A near-term priority is \textbf{portability} for surrogate models beyond the training environment across geometry families and hardware. This will be achieved by enriching the input with quantitative geometric descriptors~\cite{Foteinos_JPDC_2014, tsolakis2021, tsolakis2024} and using transfer learning to adapt with minimal new data. A second priority is composition, pairing surrogates with learned controllers to form an end-to-end AI-assisted workflow that partitions the domain, balances work, refines adaptively, and tunes runtime parameters under a single optimization loop. If realized, these advances would shift parallel meshing from a hand-tuned craft to a data-driven and reusable service embedded in large-scale simulation pipelines.

N. Chrisochoides acknowledges support from The Richard T. Cheng Endowment and thanks Kevin Garner and Min Dong for assistance with data curation and statistical analysis. Generative AI tools (Gemini, Grammarly) were used for language editing; the author reviewed and assumes full responsibility for the final content.

\sectionauthor{N. Chrisochoides}

\section{Scripting Automation}

%
%

Preparing CAD models for simulation frequently relies on scripting as the standard practice. While some interactive operations can be performed ad hoc, most engineering simulations demand reproducibility and careful documentation. Scripts, often in the form of journal files containing Python or native CAD commands, serve this purpose by archiving the exact sequence of geometry operations applied to a model. This ensures that the process of simplifying and idealizing geometry can be repeated, audited, or adapted for future analyses.

Developing these scripts is rarely straightforward. Analysts typically discover effective command sequences through interactive trial and error: applying operations, attempting to mesh, revising geometry, and repeating until a successful workflow emerges. The difficulty arises because many geometry preparation problems require ordered sequences of operations rather than single isolated fixes. Defeaturing, for example, may involve multiple removals in a particular order. Block decomposition for hexahedral meshing requires long structured command sequences to partition geometry into meshable sub-blocks. Thin volume reduction~\cite{owen2024cad} demands careful transformations to collapse narrow solids into sheet bodies without breaking topology. Across all of these applications, the analyst’s final output is a script that embodies this iterative, experience-driven process.

Traditional approaches to scripting rely heavily on hand-crafted heuristics and expert intuition. Procedures are often encoded feature by feature, with analysts building libraries of macros or journal files that address narrow classes of problems. While effective in specific contexts, these methods are labor-intensive, brittle, and difficult to generalize across diverse geometries or simulation targets.

Reinforcement learning (RL) provides a natural way to address this challenge by framing script generation as a sequential decision-making problem. By treating CAD software as the environment, geometry operations as actions, and meshing or analysis outcomes as reward signals, RL agents can learn not only which operations to apply but also the order in which to apply them. In practice, the environment is typically provided through an API to an existing geometry kernel, such as Cubit \cite{CubitUserDoc1706_2025} or Open Cascade~\cite{OpenCascadeWebsite}, which exposes modeling operations to the agent. Training proceeds through trial-and-error interaction and can be initialized from existing analyst scripts, while reward functions encode priorities such as meshability, robustness, and fidelity to the target physics. In this way, RL has the potential to automate the generation of scripts that prepare CAD models for simulation, reducing manual effort and improving both efficiency and reproducibility.

Recent work has demonstrated the viability of this approach in several application areas. In the context of block decomposition for hexahedral meshing, DiPrete et al.~\cite{diprete2024reinforcement} and Zhang et al.~\cite{zhang2025reinforcement} designed RL agents whose actions correspond to decomposition operations. The agents learn to select reference entities and cutting operations that subdivide complex geometries into meshable sub-blocks. The reward functions in these studies were crafted to encourage the formation of well-shaped sub-blocks with favorable surface and dihedral angles, while penalizing irregular cuts that degrade mesh quality. To support effective training, Zhang et al.~\cite{zhang2025reinforcement} also introduced a graph-based representation of CAD models. This representation maps B-rep entities into graph structures that capture geometry and topology, enabling the neural network to reason about relationships that matter for decomposition. Using this setup, they reported hex-meshable volume ratios exceeding 95\% on their benchmark problems, a notable milestone for automated decomposition.

Another example is the reduction of thin volumes in CAD models, a task that often requires ordered command sequences to collapse narrow solids into topologically valid sheet bodies. Owen et al.\ \cite{owen2024cad} applied RL to this problem by defining actions that select thin volumes and determine appropriate sheet body placements. Their framework learns a policy to drive these ordered operations within a CAD kernel, using per-volume agents and a connectivity graph to coordinate decisions across the model. This approach preserves model topology while simplifying the geometry, and it is now being extended to automate defeaturing operations that improve tetrahedral mesh quality and reduce element counts. The long-term goal of this line of work is to provide analysts with automated script generation that mirrors expert decision-making, thereby reducing manual effort and accelerating the overall modeling-to-simulation workflow.

A key practical question is how to evaluate generated scripts in ways that reflect engineering needs. Beyond task completion, relevant criteria include meshing success rates, mesh quality distributions, element count and resolution relative to target physics, robustness under small geometric variations, and the degree to which scripts preserve functional and interface features. Related to this, real workflows impose hard constraints, for example preserving watertightness, respecting tolerances, and avoiding topology changes that invalidate boundary conditions. Incorporating these constraints into the action space and reward design is often necessary for RL policies to be deployable in production settings.

Although these early demonstrations show considerable promise, the maturity of RL methods for geometry preparation is still at an early stage. Most studies remain proofs of concept, focusing on constrained problem classes such as block decomposition or thin volume reduction. These results confirm that RL can, in principle, learn sequences of CAD operations that prepare models for simulation, but the methods have not yet been validated across the wide range of geometries encountered in real engineering practice. One notable exception is the thin volume reduction work of Owen et al.~\cite{owen2024cad}, which has been integrated into Cubit \cite{CubitUserDoc1706_2025} and is already providing useful capability. Even here, broader training data, clearer success criteria, and wider analyst usage are needed before the approach can be considered mature. More broadly, generalization remains challenging due to CAD heterogeneity, kernel and version differences, tolerance-sensitive predicates, and long-horizon workflows where rewards are sparse.

The central challenge is that the outcome of interest is a script or procedure, but RL can only learn such procedures if the geometry is represented in a way that makes the effects of operations visible to the learning algorithm. In other words, the model must be able to detect differences in the geometry as it changes step by step, so that it can learn which sequences of operations lead to improved outcomes. Designing representations that expose these incremental changes is therefore critical.

Several strategies are being explored to address this issue. One approach is to convert B-rep entities into structured formats that can be processed by neural networks, such as sequence-based encodings~\cite{zhang2024brep2seq,zhang2024brepmfr}. Another is to use graph-based representations, where nodes and edges capture both geometry and topology, and graph neural networks (GNNs) are applied to process this information~\cite{wu2020comprehensive}. Graph-based methods are particularly promising because they allow the RL agent to reason about how operations modify geometry and topology in ways directly relevant to meshing and analysis. By combining these representation strategies with reinforcement learning, it may be possible to train agents that not only reproduce expert scripts but also discover new and more efficient procedures for geometry preparation. Looking ahead, these RL capabilities are also complementary to language-model-based scripting, where natural-language intent can propose candidate procedures and RL can refine them through closed-loop interaction with the geometry kernel.

\sectionauthor{N. Winovich}

\section{Language-Driven Scripting}

%

While reinforcement learning offers a framework for discovering sequences of CAD operations through trial and error interaction with geometry software, large language models (LLMs) \cite{zhao2023survey} address a complementary aspect of the problem: the automation of CAD scripting through direct code synthesis. RL learns strategies over many iterations; LLMs translate natural language prompts into executable scripts. Within geometry preparation and meshing, these approaches are mutually reinforcing. RL develops procedural strategies, while LLMs provide an interface for capturing, reusing, and deploying these strategies in reproducible form.

The motivation for LLM-driven scripting stems from the same challenges that have long motivated automation: CAD workflows require careful, reproducible documentation of geometry operations. Analysts traditionally create scripts manually, either through tedious trial and error in a graphical interface or by directly writing commands. While effective, this practice is slow, error prone, and demands specialized expertise. LLMs lower these barriers by allowing analysts to describe goals in natural language. From these descriptions, the model generates CAD commands that reproduce geometry, apply defeaturing, initiate meshing procedures, or prepare models for analysis. The user remains in control because the generated code is transparent and can be inspected, refined, or overridden as needed.

Three themes characterize current applications of LLMs to CAD scripting. First, direct code generation uses fine-tuned models and \emph{database retrieval-augmentation} frameworks to translate text into syntactically correct and semantically meaningful CAD code~\cite{R1,R4,R13}. These approaches often employ structured function calling or specialized annotations that capture the semantics of operations such as extrusions, fillets, and lofts~\cite{N1}. Iterative refinement, whether automated or guided by user feedback, improves robustness by detecting errors and resynthesizing code~\cite{R5,R6,R14}.

Second, multimodal input extends beyond text by combining language with sketches, images, or point clouds, enabling workflows in which a designer supplies partial visual guidance that the model converts into complete parametric scripts. Examples include models trained on large datasets of CAD images paired with code~\cite{R13,R18,R26}, as well as fine-tuned multimodal LLMs that directly edit boundary representations from text prompts while preserving geometric integrity~\cite{N2}. These methods move toward richer interactions where LLMs \emph{act as design assistants} that interpret intent across sketches, images, and point clouds and translate it into complete parametric scripts.

Third, verification and correction are integral because CAD scripting demands high fidelity. Many workflows compare generated geometry against target specifications and iteratively correct deviations \cite{R5}. Others use auxiliary vision--language models to check code execution and flag errors for resynthesis. Menon et al.\ present a multi-agent design assistant (MADA) \cite{Menon2026MultiAgentISC} that exemplifies this theme by grounding command synthesis in documentation retrieval, checking mesh metrics, and looping through scheduler-managed runs to converge on target quantities of interest.

In MADA, the agentic workflow is organized around complementary roles. A Geometry Agent creates and edits CAD, invokes defeaturing utilities, and generates meshes. A Job Agent manages scheduler interaction to launch and monitor ensembles at scale. An Inverse Design Agent explores parameters using simulation outputs or surrogate models. This division mirrors how engineering teams split modeling, meshing, and analysis, and it illustrates how LLM-driven scripting and RL-style search can be orchestrated within a single closed loop~\cite{Menon2026MultiAgentISC}. Related work extends language-driven workflows beyond CAD code to mesh artifacts themselves. LLaMA-Mesh~\cite{wang2024llamamesh} encodes mesh geometry and connectivity in a text-compatible form, enabling instruction-driven mesh generation and editing that can support similar verification and refinement loops in mesh-centered pipelines.

Together, these advances show that LLMs complement the procedural learning strengths of RL by providing a unifying interface that connects geometry representations, meshing technologies, and AI-driven methods discussed throughout this paper. Where RL agents may learn how to optimally defeature a model or decompose it for hexahedral meshing, an LLM can package such procedures into user-accessible scripts. Representations such as graph neural networks or B-rep encodings, originally motivated by the need to support RL training~\cite{zhang2024brep2seq,zhang2024brepmfr,wu2020comprehensive}, can also support LLMs by enriching the context used during script generation.

Looking forward, the convergence of these technologies points toward AI agents \cite{fang2025selfevolving} for CAD and meshing. Agents combine language understanding with iterative reasoning, tool usage, and memory. In geometry preparation, such agents could \emph{autonomously} call geometry kernels, verify outputs, and refine operations while engaging users through natural language. This unifies the strengths of both paradigms: RL provides procedural building blocks, LLMs make those procedures accessible through high-level prompts, and agent architectures orchestrate the process in a closed loop. Experience with the assistant suggests that the pattern scales to realistic CAD-to-mesh-to-solver use cases and to HPC settings \cite{Menon2026MultiAgentISC}.

\textbf{Expanding domain-specific and multimodal datasets:}
The lack of large-scale, high-quality CAD-specific and multimodal datasets limits generalization. Priority should be placed on broader datasets that capture complex geometries and industrial contexts \cite{R2,R9,R18}.

\textbf{Improving spatial reasoning and geometric fidelity:}
While LLMs can produce syntactically valid CAD scripts, precision remains challenging for tightly constrained designs. Solver-aided methods, neurosymbolic frameworks, or stronger spatial localization could improve fidelity \cite{R10,R23,R25}.

\textbf{Extending beyond geometry to full design-to-manufacturing pipelines:}
Research should incorporate design evaluation, optimization, feature tagging for simulation and FEA, and downstream manufacturing integration to enable end-to-end AI-assisted design systems \cite{R12,R19,R20}.

In addressing these limitations, researchers can accelerate the evolution of LLMs to achieve a transformative shift in the digital design cycle. By automating the translation of natural language into precise and executable CAD code, these systems promise to accelerate the design process while lowering the threshold for broad adoption.

\sectionauthor{N. Brown}
\section{Conclusion}


This review set out with a simple premise: that artificial intelligence can serve as a \emph{facilitating technology} for geometry preparation and meshing, complementing rather than replacing decades of foundational research in kernels, algorithms, and structured workflows. The survey has shown how that premise now manifests across the CAD-to-mesh pipeline. We examined methods for part classification and segmentation that automate the recognition of features within large assemblies, predictors that anticipate mesh quality before elements are generated, and AI-driven approaches to defeaturing that recommend cleanup operations. We reviewed early demonstrations of reinforcement learning for unstructured and block-structured meshing, new strategies for volumetric parameterizations, and surrogate modeling that accelerates parallel meshing. Finally, we highlighted how reinforcement learning and large language models can automate scripting, capturing analyst expertise in reproducible form.


Together, these advances show that AI is already reshaping how models are prepared for simulation, even if most methods remain at an early stage of maturity. This survey has highlighted representative technologies, illustrated how learning can extend established meshing practice, and pointed toward emerging opportunities across the pipeline. At the same time, it provides only a snapshot of a rapidly evolving field: important pieces are still missing, and the approaches described here will inevitably be refined, extended, and in some cases surpassed as new models, datasets, and workflows emerge.

Across these topics, several cross-cutting themes emerge. Most approaches face common bottlenecks: the lack of large curated datasets, standardized benchmarks, and robust geometry representations that can transfer across diverse CAD sources. The maturity of methods also varies widely. Classical machine learning for classification and mesh quality prediction has already entered production software such as Cubit \cite{CubitUserDoc1706_2025}, delivering measurable efficiency gains. In contrast, reinforcement learning for meshing and volumetric parameterization remain at the prototype stage, while segmentation and defeaturing tools are intermediate—promising in research settings but not yet broadly deployed. The most promising direction is integration: connecting classification, prediction, and defeaturing into closed-loop workflows; combining supervised and reinforcement learning; and leveraging large language models for scripting and natural interfaces. Progress in the next five years will depend less on isolated algorithmic advances than on building shared datasets, evaluation standards, and hybrid pipelines that unify AI with established meshing practice.

In this spirit, the review is not intended as a final word but as a springboard. We hope it serves as both a resource for practitioners seeking to incorporate AI into their workflows and a guidepost for researchers advancing the next generation of tools. The open challenges ahead, including richer data curation, standardized evaluation, B-rep-native learning, uncertainty-aware decision-making, and orchestration across the full simulation pipeline, offer fertile ground for continued exploration. By capturing current progress while pointing toward what lies ahead, we aim to support innovation across the community and invite future contributions that will ultimately advance and surpass the capabilities described here, bringing us closer to overcoming the long-standing CAD-to-mesh bottleneck in simulation.

\bmhead{Acknowledgments}

Chrisochoides is funded by The Richard T. Cheng Endowment. He thanks Dr. Kevin Garner and Min Dong assisted in curating the data for both shared memory (NUMA) and distributed memory parallel mesh generation runs at CRTC and statistics from the wait-time of those runs at ODU's HPC Clusters, respectively. 

The authors used ChatGPT, Gemini and Gramerly for language editing and readability improvements. All authors reviewed the manuscript and take full responsibility for the final content.

\bibliography{references}

\end{document}